\pdfoutput=1%
\documentclass[a4paper,british,floatfix,pdftex,superscriptaddress,twoside,%
aps,prl,%
longbibliography,% only for revtex4-1 for arXiv
reprint,twocolumn%,final% add final to see real layout
]{revtex4-1}%
\usepackage{amsmath,amssymb}
\usepackage[T1]{fontenc}
\usepackage{graphicx}%
\usepackage[utf8]{inputenc}
\usepackage{microtype}
\usepackage{xspace}
\usepackage{hyperref, hypernat}
\usepackage[notodos]{CMImacros}

\graphicspath{{./figures/}} % define figure directory
\hypersetup{citebordercolor=yellow,linkbordercolor=lime,urlbordercolor=green}

\newcommand*{\diff}{\mathop{}\!\mathrm{d}}%

\newcommand*{\refsupp}[1]{\hyperref[#1]{Supplementary Figure~\ref{#1}\xspace}}
\newcommand*{\subrefsupp}[2]{\hyperref[#1]{Supplementary Figure~\ref{#1}~#2\xspace}}
\newcommand*{\suppinfone}{\hyperref[sec:numerical_simulations]{Appendix~A}\xspace}%
\newcommand*{\suppinftwo}{\hyperref[sec:semiclass_simulations]{Appendix~B}\xspace}%

\newcommand{\cfeldesy}{\affiliation{Center for Free-Electron Laser Science, Deutsches
      Elektronen-Synchrotron DESY, Notkestraße 85, 22607 Hamburg, Germany}}%
\newcommand{\cfelmpsd}{\affiliation{Max Planck Institute for the Structure and Dynamics of Matter
      and Center for Free-Electron Laser Science, 22761 Hamburg, Germany}}%
\newcommand{\simons}{\affiliation{Center for Computational Quantum Physics (CCQ), The Flatiron
      Institute, 162 Fifth Avenue, New York NY 10010}}%
\newcommand{\uhhchem}{\affiliation{Department of Chemistry, Universität Hamburg,
      Martin-Luther-King-Platz 6, 20146 Hamburg, Germany}}%
\newcommand{\uhhcui}{\affiliation{The Hamburg Center for Ultrafast Imaging, Universität Hamburg,
      Luruper Chaussee 149, 22761 Hamburg, Germany}}%
\newcommand{\uhhphys}{\affiliation{Department of Physics, Universität Hamburg, Luruper Chaussee 149,
      22761 Hamburg, Germany}}%
\newcommand{\upaphys}{\affiliation{Dipartimento di Fisica e Chimica, Universitá degli Studi di
      Palermo, Via Archirafi 36, I-90123, Palermo, Italy}}%

\begin{document}
\title{Setting the photoelectron clock through molecular alignment}%
\author{Andrea Trabattoni}\cfeldesy\uhhcui%
\author{Joss Wiese}\cfeldesy\uhhchem%
\author{Umberto De Giovannini}\cfelmpsd\upaphys%
\author{Jean-Fran\c{c}ois Olivieri}\cfeldesy%
\author{Terry~Mullins}\cfeldesy%
\author{Jolijn~Onvlee}\cfeldesy%
\author{Sang-Kil~Son}\cfeldesy\uhhcui%
\author{Biagio Frusteri}\upaphys%
\author{Angel Rubio}\cfelmpsd\simons\uhhphys%
\author{Sebastian Trippel}\cfeldesy\uhhcui%
\author{Jochen Küpper}\cfeldesy\uhhcui\uhhchem\uhhphys%
\date{\today}%
\begin{abstract}\noindent%
   The interaction of strong laser fields with matter intrinsically provides powerful tools to image
   transient dynamics with an extremely high spatiotemporal resolution. Here, we study strong-field
   ionisation of laser-aligned molecules and show a full real-time picture of the photoelectron
   dynamics in the combined action of the laser field and the molecular interaction. We demonstrate
   that the molecule has a dramatic impact on the overall strong-field dynamics: it sets the clock
   for the emission of electrons with a given rescattering kinetic energy. This result represents a
   benchmark for the seminal statements of molecular-frame strong-field physics and has strong
   impact on the interpretation of self-diffraction experiments. Furthermore, the resulting encoding
   of the time-energy relation in molecular-frame photoelectron momentum distributions shows the way
   of probing the molecular potential in real-time and accessing a deeper understanding of electron
   transport during strong-field interactions.
\end{abstract}
\maketitle%

\section*{Introduction}
In the prototypical strong-field interaction, an intense driving field extracts a valence electron
from the target through tunnel ionisation, accelerates the free electron in vacuum, and eventually
drives it back to the parent ion, predominantly resulting in rescattering or radiative
recombination~\cite{Corkum:PRL71:1994, Corkum:ARPC48:387}. The radiative recombination results in
the emission of high-energy photons by high-harmonic generation~\cite{Corkum:PRL71:1994} and this is
a powerful tool to investigate the electronic structure with attosecond temporal
resolution~\cite{Calegari:Science346:336, Lepine:NatPhoton8:195, Kraus:Science350:790}.
Alternatively, the rescattered portion of this electron wavepacket is exploited in laser-induced
electron diffraction (LIED)~\cite{Corkum:NatPhys3:381} experiments as a coherent diffraction pattern
of the molecular target, potentially providing time-dependent images of the molecule at
sub-femtosecond and few-picometer resolution. Recently, corresponding experimental results for the
structure or dynamics of small or highly-symmetric molecules were
obtained~\cite{Blaga:Nature483:194, Pullen:NatComm6:7262, Wolter:Science354:308,
   Walt:NatComm8:15651, Fuest:PRL122:053002, Amini:PNAS116:8173}. At the same time, the initial
conditions of the strong-field interaction have attracted much attention for capturing the intrinsic
nature of strong-field physics.

While pioneering attosecond experiments and molecular-frame measurements revealed non-trivial
spatiotemporal features in electron tunnelling~\cite{Eckle:Science322:1525, Meckel:NatPhys10:594},
these initial conditions are still generally considered a weak perturbation in strong-field physics.
All the results obtained in LIED experiments, for example, are interpreted in the framework of the
strong-field approximation, where the electron is considered to be born in the continuum with a
negligible initial momentum and to propagate as a plane wave~\cite{Chen:PRA79:033409}. Furthermore,
the post-ionisation dynamics before rescattering are assumed to be fully driven by the laser field,
by neglecting, for example, the Coulomb interaction with the ionised molecule.

Common strategies to analyse photoelectron momentum distributions rely on the quantitative
rescattering theory (QRS)~\cite{Chen:PRA79:033409}, where angular dependence in the final
photoelectron wavepacket is introduced solely through rescattering. Within this approach,
diffraction patterns were analysed utilising the angular~\cite{Blaga:Nature483:194,
   Pullen:NatComm6:7262} or radial~\cite{Xu:NatComm5:4635} photoelectron distribution. However, the
relevance of the ionised molecular orbital in the rescattered photoelectrons is still under
discussion~\cite{Schell:SciAdv4:eaap8148}. So far, this was included by an overall weighting factor
in the rescattering probability~\cite{Lein:JPB36:L155, Niikura:Nature417:917}, or as a spatial phase
or an angular feature in the rescattering electron wavepacket~\cite{Busuladzic:PRL100:203003,
   Meckel:NatPhys10:594}. Recently, the influence of molecular alignment on molecular structure
retrieval was discussed~\cite{Xu:NatComm5:4635, Pullen:NatComm7:11922}. However, general predictions
are still extremely challenging with new models appearing~\cite{Suarez:PRA94:043423,
   Liu:PRL116:163004}.
\begin{figure*}[t]
   \centering%
   \includegraphics[width=0.66\linewidth]{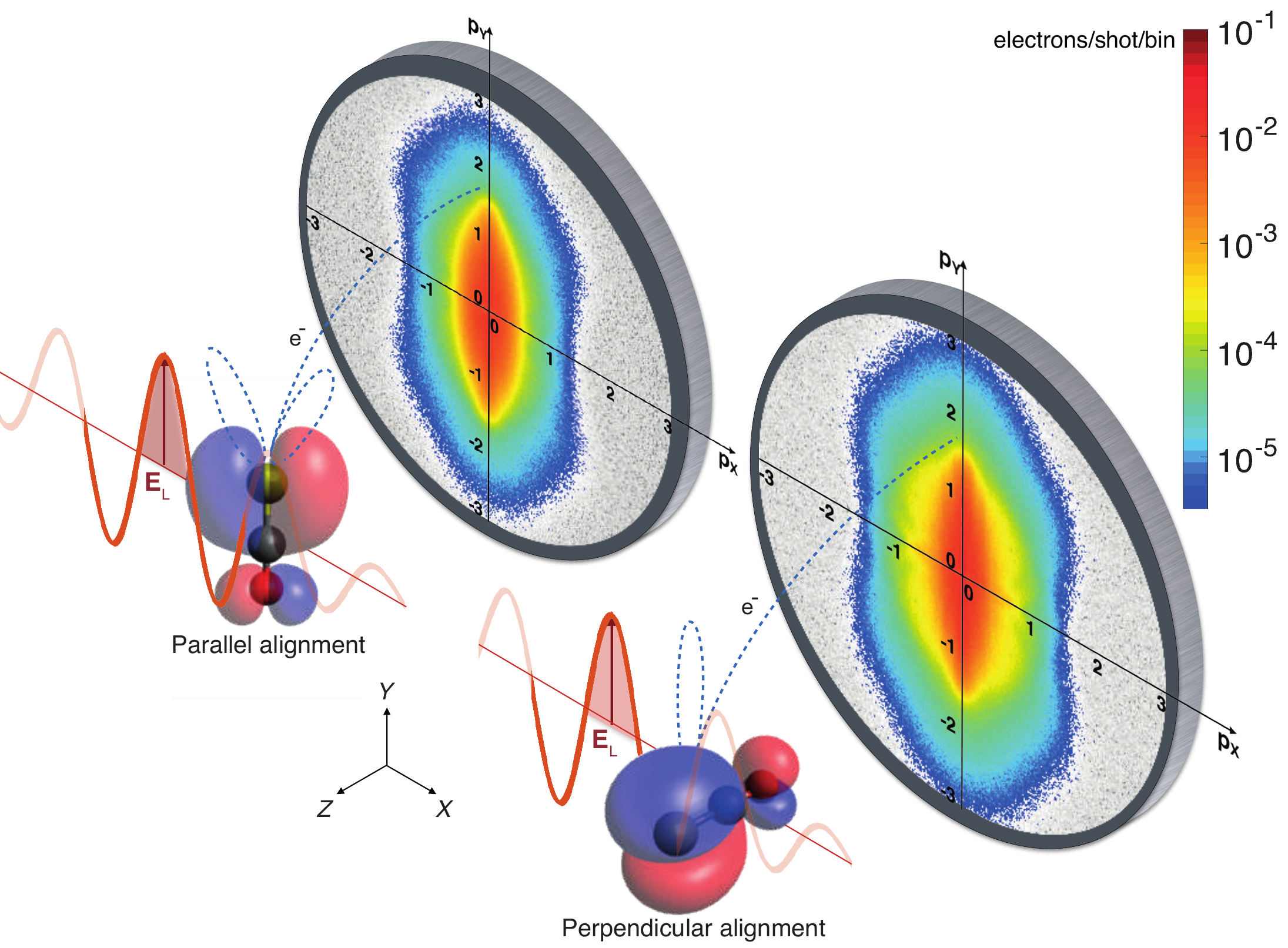}% seems like a standard figure width
   \caption{Sketch of the experimental arrangement. OCS molecules (O in red, C in black, S in
      yellow) were aligned in the laboratory frame, parallel and perpendicular to the $Y$ axis. The
      ionising laser electric field ($\mathbf{E}_\text{L}$) was linearly polarised along the $Y$
      axis and the detection was in the $XY$ plane. The molecular-frame angle-resolved photoelectron
      spectra were projected onto a 2D detector in a velocity-map-imaging spectrometer. The
      alignment-dependent photoelectron trajectories are pictorially shown (blue dashed lines), as
      well as the corresponding shape of the ionising orbital (blue and red lobes). The spectra are
      displayed on a logarithmic intensity scale in units of electrons/shot/bin.}
   \label{fig:1}
\end{figure*}

Here, we experimentally and computationally study molecular-frame photoelectron spectroscopy from
strongly aligned molecules in order to investigate the relation between the molecular frame and the
strong-field-induced ultrafast electron dynamics. We demonstrate that and how the molecular frame
governs the rescattering time for the photoelectron and, consequently, its final kinetic energy.

\section*{Results}
\subsection{Experimental approach}
\autoref{fig:1} depicts the experiment. An ensemble of carbonyl sulphide (OCS) molecules all in the
rovibronic ground state~\cite{Chang:IRPC34:557} was adiabatically aligned in the laboratory frame,
with $\cost=0.9$, by using a linearly polarised, 500~ps laser pulse, centred at
800~nm~\cite{Trippel:MP111:1738, Trippel:PRA89:051401R}, with a peak intensity
$I=3\times10^{11}~\text{W\,cm}^{-2}$. The molecules were aligned in two different configurations,
shown in \autoref{fig:1}, with the molecular axis along the $Y$ and $Z$ axes, named parallel and
perpendicular alignment, respectively. A second laser pulse, centred at $1300$~nm, with a duration
of $65$~fs, and a peak intensity $I=8\times10^{13}~\text{W\,cm}^{-2}$, was used to singly ionise the
OCS molecules. For this intensity the ponderomotive energy of the laser field is $\Up{}\approx13$~eV
and the ionisation occurred in the tunnelling regime. The electric field of the ionising laser
pulse, $\mathbf{E}_\text{L}$ in \autoref{fig:1}, was linearly polarised along the $Y$ axis
(ellipticity $\epsilon=I_Z/I_Y<0.005$). The produced molecular-frame angle-resolved photoelectron
spectra (MF-ARPES) were recorded in a velocity map imaging (VMI)
spectrometer~\cite{Eppink:RSI68:3477} with its detector parallel to the $XY$ plane. It is important
to note that the de Broglie wavelength of rescattering electrons in the experiment was larger than
$200$~pm. In this regime no diffraction feature is expected to appear in the photoelectron
distributions~\cite{Xu:NatComm5:4635}.

\subsection{Photoelectron-momentum distributions}
\autoref{fig:1} shows the MF-ARPES for parallel (left) and perpendicular (right) alignment. The two
distributions show several differences. The spectrum for parallel alignment has a larger width at
small transverse momenta, $p_X<0.5~\au$ (atomic units), while the spectrum for perpendicular
alignment shows a number of angular features for transverse momenta $p_X$ between $0.5~\au$ and
$1~\au$.
\begin{figure*}
   \includegraphics[width=\linewidth]{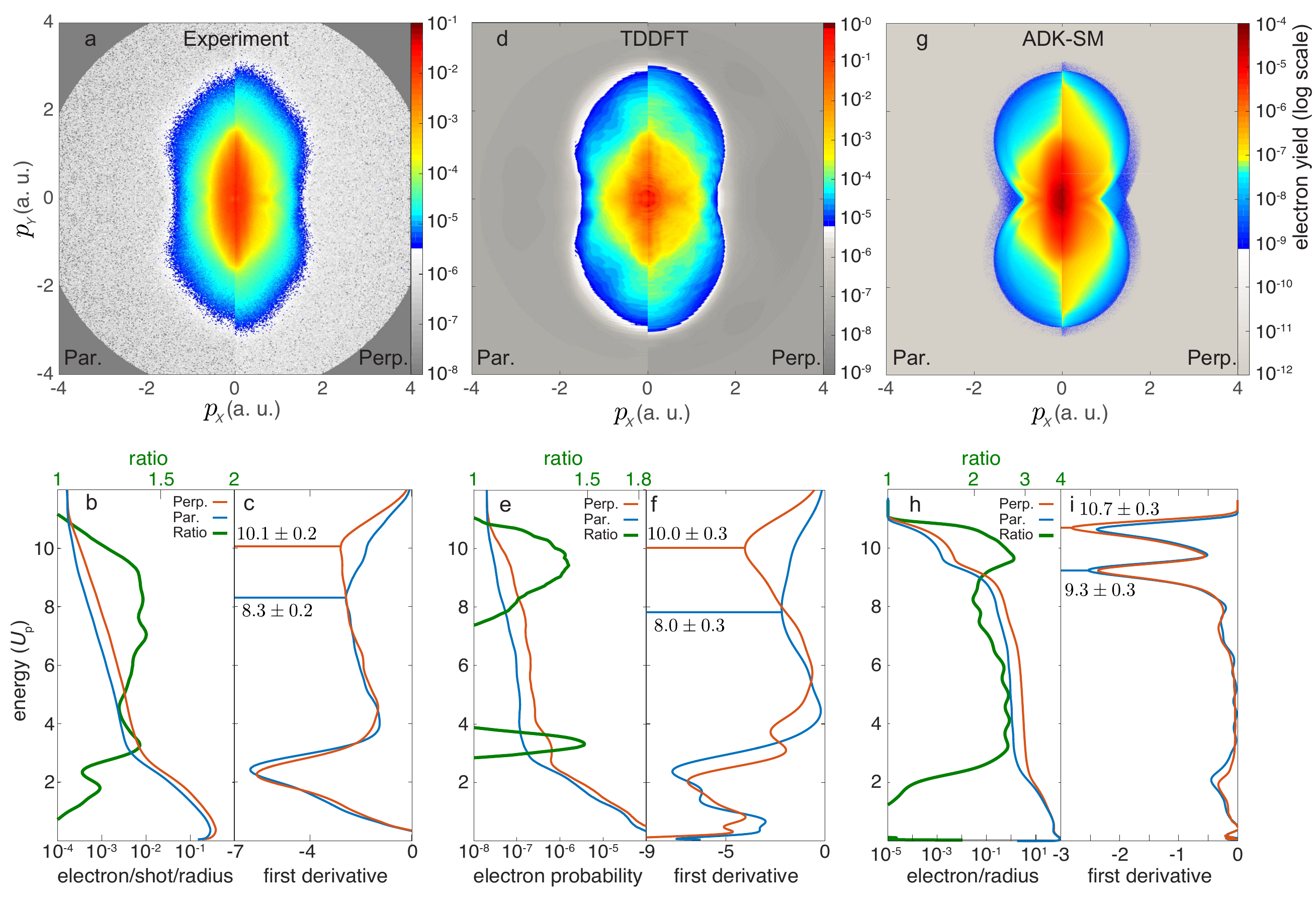}%
   \caption{Molecular-frame angle-resolved photoelectron spectra of OCS. These data were obtained
      \textbf{a,~b,~c} experimentally and computationally from \textbf{d,~e,~f} TDDFT and
      \textbf{g,~h,~i} ADK-SM calculations. \textbf{a,~d,~g}: Split graphical representation as a
      comparison of the photoelectron distributions for parallel and perpendicular alignment for the
      experimental and computational results, respectively. \textbf{b,~e,~h}:~Corresponding
      projected energy distributions of photoelectrons along the $Y$ axis, angularly integrated
      within a cone of $\degree{\pm20}$, as well as the ratio of the integral-normalised
      perpendicular and parallel distributions, on logarithmic scales. Energies are reported in
      units of the ponderomotive energy \Up{}. \textbf{c,~f,~i}: First derivatives of the
      photoelectron-energy distributions to evaluate the high-energy cutoff for the two
      molecular-alignment cases. All TDDFT computational results were obtained by averaging over
      different laser-molecule orientations according to the experimental alignment distributions
      and by adding a constant to account for the experimental background level. The ADK-SM results
      refer to a single laser intensity and perfect alignment for both cases. See \suppinftwo for
      details.}
   \label{fig:2}
\end{figure*}
These angular structures, which are much weaker in the spectrum for parallel alignment, could be
identified as forward-rescattering features~\cite{Becker:JPB48:151001}. Focussing the attention on
large longitudinal momenta $p_Y$, the counts for parallel alignment drop around $2.5$~\au. In the
case of perpendicular alignment, however, the spectrum extends to larger momenta, showing an
appreciable amount of counts at $p_Y=3~\au$. Following the strong-field approximation, the hard
cutoff of photoelectron momentum is expected to only depend on the properties of the laser
field~\cite{Paulus:JPB27:L703}. Experimentally, this quantity is hard to measure. Thus, the
turning-point of the signal drop, \ie, the minimum of the first derivative, at large longitudinal
momenta is used instead. In the following, we use the term cutoff in the latter sense. Surprisingly,
in the current study we found a clear dependence of the cutoff on the molecular frame.

\autoref[a]{fig:2} shows a close comparison of the two experimental distributions for the complete
range of $p_X$ and $p_Y$, between $0$ and $4$~\au. Here, the spectra were split along the $Y$ axis
and the spectrum from parallel alignment is shown on the left and the one from perpendicular
alignment on the right. Now, the differences at small momenta as well as at the cutoff are even more
evident. To perform a quantitative analysis of the cutoffs, the momentum distributions were
angularly integrated within a cone of $\pm\degree{20}$ with respect to the longitudinal axis ($Y$)
and converted to an energy scale. In \subautoref{fig:2}{b} the resulting photoelectron spectra are
shown for parallel (blue) and perpendicular (red) alignment, with energies in units of \Up{}. The
perpendicular/parallel ratio of the two area-normalised spectra (green) shows a predominance of
photoelectrons for perpendicular alignment in the energy range between \Up{2} and \Up{10}, where the
distribution is dominated by rescattered electrons~\cite{Becker:AAMOP48:35}. Furthermore, the ratio
increases with energy, reaching the maximum around the cutoff. To evaluate the cutoffs, the first
derivative of the energy distributions are shown in \autoref[c]{fig:2} and their minima were used to
find the edges of the distributions, which allowed us to analyse the cutoff region. The first
minimum represents the drop of direct electrons~\cite{Becker:AAMOP48:35} and it was around \Up{2}
for both alignment cases. This excluded any significant alignment-dependent direct-electron-cutoff
enhancement~\cite{Smeenk:PRL112:253001}. Surprisingly, the second minimum behaves differently for
the two alignments. While it is located around \Up{10} for perpendicular alignment, as expected from
the well established above-threshold ionisation theory~\cite{Paulus:JPB27:L703}, the cutoff is
shifted down to a value around \Up{8.5} for parallel alignment.

\subsection*{Quantum-mechanical model \\ of the electron dynamics}
To unravel the experimental observations, state-of-the-art calculations were performed using both,
time-dependent density-functional theory (TDDFT)~\cite{Marques:TDDFT:2011} and a semiclassical
molecular trajectory simulation setup. Using TDDFT, the MF-ARPES probability was calculated by
simulating the complete dynamics of the many-body ionisation process in real-time and real-space
with the tSURFF method~\cite{Tao:NJP14:2012, Wopperer:EPJB90:2017}, see \suppinfone for details.
With this technique the spectrum was obtained by computing the entire time-dependent electron
dynamics, including many-body electron interactions, and collecting the flux of electrons through a
closed surface surrounding the molecule. \autoref[d,~e,~f]{fig:2} report the same analysis of the
numerical results as performed for the experimental data in \autoref[a,~b,~c]{fig:2}. The
simulations capture the principal features of experimental data very well. In particular,
\subautoref{fig:2}{f} shows that the calculations reproduce the experimental cutoff positions for
parallel and perpendicular alignment as well as the corresponding shift between them very well. This
result is strongly affected by the electron-electron interaction and the interplay between different
orbitals. Indeed, it is evident from the calculation that the molecule is predominantly ionised from
the highest-occupied molecular orbital (HOMO) for both alignments. In the case of parallel
alignment, nevertheless, a small contribution of HOMO-1 to the yield of high-energy rescattered
electrons is observed. When the electron-electron interaction is artificially turned off the HOMO-1
contribution becomes significant and in this scenario the reduced cutoff observed in the experiment
is not reproduced, see \refsupp{fig:edf4}. Instead, in the case of fully interacting electrons the yield
of the rescattered electrons ionised from HOMO-1 is suppressed, resulting in the really good
agreement with the experiment.

\subsection*{Semiclassical model of the electron dynamics}
Furthermore, semiclassical trajectory simulations based on the Ammosov-Delone-Kra\v{\i}nov (ADK)
tunnelling theory~\cite{Ammosov:SVJETP64:1191} in conjunction with a simple man propagation
(SM)~\cite{Gallagher:PRL61:2304, Corkum:PRL62:1259} were conducted in order to track the
molecular-frame electron dynamics during the strong-field interaction~\cite{Wiese:PRX2020:inprep},
see \suppinftwo for details. Based on the TDDFT analysis of the different molecular orbitals
contributing to the photoelectron dynamics, the ionisation was assumed to occur solely from HOMO. In
the underlying model, the initial phase-space distribution of the electron wavepacket in the
continuum at birth was described by the quasistatic ADK tunnelling theory, and the nodal structure
of the HOMO was accounted for as imprint onto this initial momentum distribution. Post-ionisation
dynamics of the electron wavepacket were evaluated in the combined interaction with the laser
electric field and the cation's Coulomb field modelled as a point charge. To evaluate the accuracy
of this semiclassical description, the resulting MF-ARPES for parallel and perpendicular alignment
were calculated and analysed, see \autoref[g,~h,~i]{fig:2}, and they show a really good agreement,
both, with the experimental data and the full TDDFT calculation, reproducing the main features and
cutoffs observed in the experiment very well. In particular, as seen by the local maximum around
\Up{10} of the ratio of the two alignment cases, \autoref[h]{fig:2}, this semiclassical model
captures the enhanced yield of high-energy rescattered electrons for perpendicular alignment with
respect to parallel alignment. This result is corroborated by the enhanced cutoff around \Up{10.7}
for perpendicular alignment in \autoref[i]{fig:2}, although a smaller yield at this energy is
present also for parallel alignment. In addition, a relevant minimum appears around \Up{9.3} for
both alignments. These features of ADK-SM, together with the pronounced yield along the centreline
of \autoref[g]{fig:2}, are known to be mainly due to Coulomb focussing~\cite{Danek:JPB51:114001},
\ie, the dynamics of a continuum electron wavepacket being focussed along a perfectly linear laser
polarisation axis. The relevance of this effect is discussed further below.
\begin{figure*}
   \includegraphics[width=\linewidth]{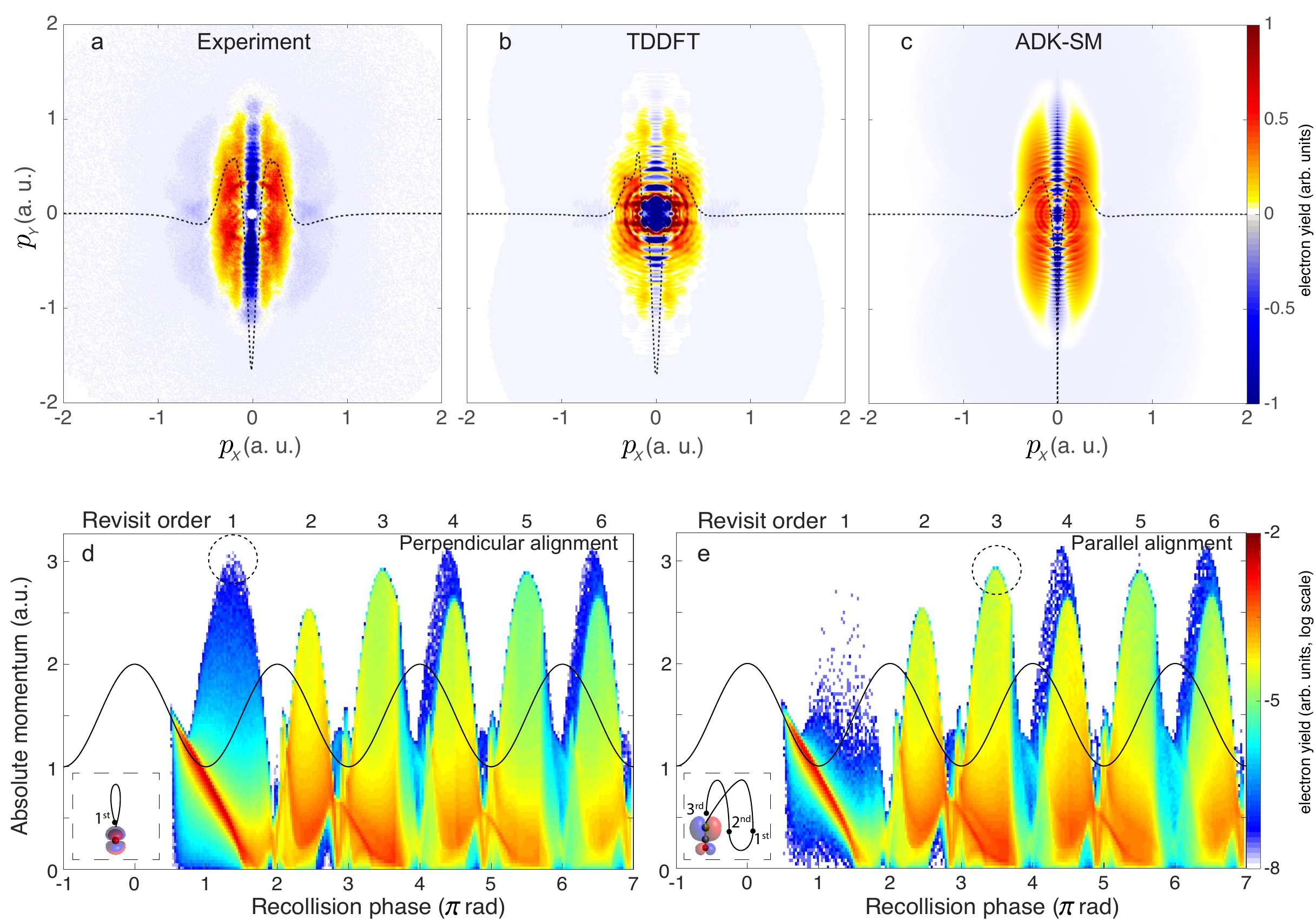}
   \caption{Differential momentum distributions and simulated final absolute photoelectron
      momentum. \textbf{a,~b,~c}: Differential momentum distributions (parallel-perpendicular) from
      \textbf{a} the experiment and the \textbf{b} TDDFT and \textbf{c} ADK-SM calculations. To
      estimate the difference of the transverse momentum component the signal is integrated along
      the $Y$ axis, shown by the black dashed lines. \textbf{d,~e}: Final absolute photoelectron
      momentum as a function of the recollision phase (bottom) and revisit order (top) for
      \textbf{d} perpendicular and \textbf{e} parallel alignment, calculated with ADK-SM. The colour
      scale maps the electron counts at every momentum-phase point. The dashed black circles
      highlight the largest-momentum electrons at the most probable revisit order for the two
      alignments. A distance of $r<5$~atomic units between electron and point charge is interpreted
      as a collision and only electrons with exactly one collision are shown. The solid black line
      depicts the external electric field. The insets give pictorial representations of
      molecular-frame electron trajectories, where the cardinals represent the revisit order. }
   \label{fig:3}
\end{figure*}

\subsection*{Differential analysis of the momentum distributions}
To obtain a more comprehensive picture of the alignment-dependent photoelectron dynamics and, in
particular, a glimpse at the initial electron wavepacket, we performed a differential analysis by
subtracting the photoelectron distributions of the two alignment cases from each other.
\autoref[a,~b,~c]{fig:3} show the relative normalised differences, parallel minus perpendicular, for
the experiment, the TDDFT simulations, and the ADK-SM calculations, respectively. The agreement
between experimental data and both models is excellent. Here, a strong depletion along the vertical
axis and two transversely offset broad lines of positive yield appear as main features, with a
really good agreement between experimental and both computational results. The depletion along the
centreline is due to the nodal structure of the degenerate $\Pi$ HOMO of OCS: it represents a
forbidden direction of electron ejection~\cite{Lein:JPB36:L155, Holmegaard:NatPhys6:428}. Therefore,
when the molecular axis was aligned along the polarisation axis of the strong field, the electron
preferentially acquired an initial transverse momentum $p_{0X}$ that was much larger than in the
case of perpendicular alignment, shown by the red vertical ridges in \autoref[a,~b,~c]{fig:3}.

\section*{Discussion}
The features observed in \autoref[a,~b,~c]{fig:3} show the crucial impact of the electronic
structure on the initial conditions of the electron at birth. However, a quantitative evaluation of
the initial conditions of the electron at tunnelling is challenging~\cite{Pfeiffer:PRL109:083002,
   Geng:PRL19:193001, Tian:PRL118:213201}. In general, they are defined by the tunnel-exit position
as well as by the temporal phase and the momentum acquired during the ionisation with respect to the
external field. Here, we demonstrate that the molecular potential, \ie, the combination of the
electrostatic potential and the electronic structure of the molecule, has in fact not only a primary
role in setting the initial conditions for electron emission, but that it also drives the whole
photoelectron dynamics: it defines the cutoff of rescattered electrons and it shapes the time-energy
relation for electron recollision. To investigate this, we exploited the ADK-SM calculations to
analyse the final absolute momentum acquired by the electron after photoionisation as a function of
the recollision phase. In the following discussion we will refer to revisit to describe passages of
the electron nearby the cation on a relatively large spatial scale, where momentum transfer is
relatively small and soft, whereas recollision, or rescattering, refers to the close approach of the
electron to a nucleus, with an associated large momentum transfer, \eg, in back scattering. While
multiple revisits may occur during the interaction of the electron with its cation, a recollision
event will drive the electron irreversibly away from the molecule. The resulting momentum
distributions are reported in \autoref[d,~e]{fig:3} for perpendicular and parallel alignment,
respectively. They consist of broad peaks appearing every half cycle of the electric field at phases
close to $(k+1/2)\pi,~k=1,2,3\ldots$, for which the electron collides with the molecular cation when
the laser field's vector potential is maximum. The first recollision event, \ie, the first peak in
\autoref[d]{fig:3} at a phase of $3\pi/2$, allows the electron to reach the largest momentum as
expected in the classical theory~\cite{Paulus:JPB27:L703}. This is close to the maximum asymptotic
kinetic energy, \ie, the \Up{10} cutoff. The peaks appearing later correspond to electrons that have
initially missed and then revisited the ion at later times. These subsequent rescattering events are
expected to lead to lower photoelectron energies~\cite{Paulus:JPB27:L703}. At the same time, these
multiple revisits are possible only due to the Coulomb attraction of the ionised
molecule~\cite{Paulus:JPB27:L703}. Since the current understanding and analysis of strong-field
self-diffraction experiments only consider the photoelectron recollision on the first
revisit~\cite{Blaga:Nature483:194, Amini:PNAS116:8173}, the relevance of Coulomb attraction is
usually neglected. However, our results demonstrate that it is a crucial ingredient to correctly
understand molecular-frame electron rescattering. Note that a small yield at large momentum
($>3~\au$) is visible for both alignments, \autoref[d,~e]{fig:3}, at the fourth and the sixth
revisit. These revivals, caused by Coulomb focussing, \emph{vide supra}~\cite{Danek:JPB51:114001},
are expected to vanish for imperfect linear polarisation of the laser field, as usually occurring in
any experiment. This explains why ADK-SM for parallel alignment has another cutoff around \Up{10.7},
as well as the more pronounced cutoff around \Up{9.3} in \autoref[i]{fig:2}. Due to the subtle
conditions of Coulomb focussing, this effect will not be further considered in the discussion below;
it does not contradict any of our general conclusions.

In this framework, the largest absolute momentum for perpendicular alignment comes from the first
rescattering event at a phase around $3\pi/2$, see \autoref[d]{fig:3} and its inset, which yields
the largest momentum $\ordsim3.15~\au$. This momentum corresponds to an asymptotic kinetic energy
$\ordsim\Up{10.5}$ and thus explains the experimental observation of the \Up{10} cutoff for
perpendicular alignment, see the red marker in \autoref[c]{fig:2}. This rescattering event is
attenuated by the imprinting of the nodal plane perpendicular to the molecular
axis~\cite{Schell:SciAdv4:eaap8148}, as otherwise this first peak would not only correspond to the
largest photoelectron momentum, but also to the most probable recollision event. This attenuation
for perpendicular alignment is responsible for the rescattering at the third revisit, \ie, at a
phase around $7\pi/2$, to play a major role at lower energies and for the build up of a secondary
cutoff at $\Up{\ordsim9}$, \autoref[i]{fig:2}. While this second distinct minimum is not clearly
visible in the first derivative of the experiment and the TDDFT calculations, \ie, the red curves in
\autoref[c,~f]{fig:2}, the broad shape of the minima in \autoref[c,~f]{fig:2} at high energy may be
in fact a signature of the attenuation of the scattering at first revisit and the relevant
contribution of the third revisit. In the case of parallel alignment, instead, the first
rescattering event is strongly suppressed and most of the large-momentum electrons come from the
third revisit at a phase of $7\pi/2$, depicted in the inset of \autoref[e]{fig:3}; the fifth revisit
also yields comparable momenta. As a result, the momentum cutoff is smaller, \ie, $\ordsim2.9~\au$,
corresponding to a final kinetic energy of $\ordsim\Up{9}$. This is in good agreement with the
experimentally observed reduced cutoff for parallel alignment, see the blue marker in
\autoref[c]{fig:2} and \autoref[f,~i]{fig:2}. This dynamics is mainly driven by the molecular
potential: Here, the node of the HOMO along the laser polarisation imprints an angle on the electron
emission at tunnelling. For OCS this angle was estimated to be $\ordsim\degree{30}$ with respect to
the longitudinal $Y$ axis by the TDDFT calculations. This angle prevents the electron from
rescattering at the first revisit. However, then the Coulomb attraction of the ionised molecule
forces the electron to stay in the interaction region and to recollide at later revisits. It is
important to note that the angle of emission and the rescattering at the $n$-th revisit are strongly
correlated. Indeed, larger emission angles lead to later revisits and \emph{vice versa}. As the
consequence, the photoelectron cutoff carries a clear signature of the electronic structure at
tunnelling. This angular dependence imprinted in the momentum distribution of the initial electron
wavepacket leads to the breakdown~\cite{Schell:SciAdv4:eaap8148} of the common product ansatz in
QRS~\cite{Chen:PRA79:033409}, where the initial and the rescattered parts of the wavepacket are
separated and only the latter is considered angularly dependent in the recollision frame.

Furthermore, the photoelectron cutoff in the molecular frame carries crucial time information: While
the cutoff for perpendicular alignment is strongly shaped by electrons recolliding $3/4$ of an
optical cycle after ionisation, as usually assumed, this is not true for parallel alignment: the
cutoff is dominated by electrons revisiting the molecule much later, namely one or multiple optical
cycles later. For a wavelength of $1300$~nm this corresponds, at least, to a delay of
$\ordsim4.3$~fs and it linearly increases with the wavelength. From \autoref[d,~e]{fig:3}, apart
from the aforementioned effects of Coulomb focussing, it is also evident that even-numbered revisits
yield lower kinetic energies $\Up{<8}$~\cite{Paulus:JPB27:L703}. Since the time spent by the
photoelectron before rescattering is usually exploited as the elementary delay step for
time-resolved self-diffraction experiments~\cite{Blaga:Nature483:194}, the use of this lower range
of photoelectron energy~\cite{Wolter:Science354:308, Amini:PNAS116:8173} results in any time
information being smeared out on much longer timescales. Furthermore, the analysis performed here
demonstrated that this delay step strongly depends on the molecular-frame alignment and that the
molecular potential sets a complex time-energy encoding in the electron dynamics. This
molecular-frame clock for electron recollision could clearly be exploited to disentangle the
structural dynamics with few-fs or even sub-fs temporal resolution. For instance, signals from the
first (few) revisit order(s) could be selected in the experiment with near-single-cycle (few-cycle)
laser pulses.

We demonstrated, experimentally and computationally, that the molecular frame determines the
momentum distribution of high-energy rescattered electrons in strong-field ionisation. The basic
concept of molecular-frame strong-field ionisation is captured by considering the initial conditions
imposed by the molecular potential in the dynamics of the photoelectron. Furthermore, from the
analysis of the rescattering trajectories it is evident that the molecular interaction plays a
crucial role in setting a clock for the emission and the dynamics of high-energy electrons. It
highlights that the molecular frame has a strong impact on the relation between the photoelectron
energy and the rescattering time. This finding redefines the delay step of time-resolved
self-diffraction experiments and opens up a perspective on time-resolved diffraction experiments.
These conclusions hold similarly for other observables related to electron recollision, \eg,
high-harmonic-generation spectroscopy.

Our result represents an important benchmark for any self-diffraction measurement and represents a
breakdown of the usual interpretation of LIED experiments~\cite{Blaga:Nature483:194,
   Wolter:Science354:308, Amini:PNAS116:8173}.
   We note that in such experiments mid-infrared lasers ($\lambda\approx3~\um$)
   are typically employed. Since the electron’s excursion
   length increases with increasing laser wavelengths, we expect our findings to be even more
   relevant for actual LIED experiments.

Our study highlights the molecular-frame conditions as a crucial ingredient of self-diffraction
experiments. This framework is general and can, in principle, be extended to any molecular system.
Furthermore, the molecular-frame strong-field interaction was quantitatively modelled here by a
fully-interacting-electron TDDFT calculation and, in conjunction, by a semiclassical
single-active-electron theory. We exploited TDDFT to evaluate the contribution of different
molecular orbitals to the ionisation-rescattering dynamics as a benchmark for the applicability of
the semiclassical approach. We expect this double-sided theoretical framework to become more and
more important with increasing molecular complexity, where modelling the photoelectron dynamics may
go beyond the capabilities of a single-orbital picture. In general, this also opens the perspective
to investigate electron-correlation-driven phenomena in molecular strong-field
physics~\cite{Lepine:NatPhoton8:195}. Furthermore, the earliest moments of a strong-field
interaction are intrinsically imprinted in the initial conditions of the photoelectron and in the
final energy distribution. Thus, molecular-frame strong-field-ionisation experiments, in principle,
allow one to achieve a deeper understanding of electron tunnelling, for instance, regarding the
tunnelling time, and to track the molecular potential in real-time.

\section*{Acknowledgements}
This work has been supported by the Clusters of Excellence ``Center for Ultrafast Imaging'' (CUI,
EXC~1074, ID~194651731) and ``Advanced Imaging of Matter'' (AIM, EXC~2056, ID~390715994) of the
Deutsche Forschungsgemeinschaft (DFG), by the European Research Council under the European Union's
Seventh Framework Programme (FP7/2007-2013) through the Consolidator Grant COMOTION
(ERC-Küpper-614507) and under the Horizon 2020 Research and Innovation Programme through the
Advanced Grant QSpec-NewMat (ERC-Rubio-694097), and by the Helmholtz Association Initiative and
Networking Fund. A.T.\ and J.O.\ gratefully acknowledge fellowships by the Alexander von Humboldt
Foundation.

\section*{Author contributions}
A.T., S.T., and J.K.\ conceived the experiment. A.T., S.T., J.F.O., J.W., T.M., and J.O.\ performed
the experiment. A.T.\ performed the data analysis. J.W.\ set up and performed the ADK-SM
simulations. U.D.G.\ and B.F. performed the TDDFT calculations, and together with S.-K.S.\ and A.R.\
provided theoretical support. A.T., S.T., U.D.G., J.W., A.R., and J.K.\ interpreted the results and
prepared the manuscript. All authors contributed to the discussion of the results and commented on
the manuscript.

\noindent\textbf{Correspondence} and requests for materials should be addressed to A.R.\
(\url{angel.rubio@mpsd.mpg.de}) and J.K.\ (\url{jochen.kuepper@cfel.de}).

\appendix
\section{Appendix~A: Numerical quantum-dynamics simulations}
\label{sec:numerical_simulations}
Numerical simulations of the full LIED dynamics have been performed from first-principles within the
time-dependent density functional theory (TDDFT)~\cite{Marques:TDDFT:2011} framework as implemented
in the real-space real-time Octopus code~\cite{Andrade:PCCP17:2015}. In TDDFT, the dynamics of an
interacting many-electron system is cast into the manageable problem of a fictitious
non-interacting system under the effect of a time-dependent potential such that the non-interacting
and the interacting systems have the same time-dependent density.

Since core electrons are expected to play a marginal role in the experiment we consider only valence
electrons and account for inner-shell electrons by the effect of norm-conserving Troullier-Martins
pseudopotentials. To obtain a good description of ionisation, we employed a local density
approximation (LDA) functional with the average density self-interaction correction
(ADSIC)~\cite{Legrand:JPBAMO35:2002}, which corrects the asymptotic decay and provides a first and
second ionisation energy of 11.65~eV and 15.69~eV, in good agreement with experimental
values~\cite{Wang:JESRP47:167}. During the simulations the nuclei are held fixed in the equilibrium
positions, $r_\text{C--S}=156.1~\text{pm}$ and $r_\text{C--O}=115.6~\text{pm}$.

The TDDFT equations are discretised in real-space with a cartesian grid of spacing 0.4~\au with a
cylindrical shape of radius 50~\au and length 260~\au aligned along the laser-polarisation
direction. The solution of the electron dynamics is obtained by using a discretised real-time
evolution with a time step of 0.08~\au. The calculations are performed with a 30~fs laser pulse.
Complex absorbing boundaries of varying thicknesses, 40~\au from the caps of the cylinder and 10~\au
on the radial borders, are placed at the edges of the simulation box to prevent spurious
reflections~\cite{DeGiovannini:EPJB88:2015}.

The photoelectron spectrum is calculated by collecting the flux of the photoionisation current
through a spherical surface of radius 40~\au with the tSURFF method~\cite{Tao:NJP14:2012,
   Wopperer:EPJB90:2017}. This approach gives access to the momentum resolved photoelectron
probability $I(\mathbf{p})$ from which, by integrating along the direction perpendicular to the
detector, it is possible to obtain the angular distribution of the experiment:
$I(p_X,p_Y)=\int\diff{p_Z}\,I(\mathbf{p})$.

The non-perfect molecular alignment in the laboratory frame is accounted for by sampling the
relative angle $\theta$ between the laser polarisation and the molecular axis from \degree{0} to
\degree{90} in steps of \degree{10}, as shown in \refsupp{fig:edf1}. This procedure requires a
separate simulation for each $\theta$. The photoelectron spectra for a given configuration, parallel
or perpendicular, are obtained by averaging the photoelectron distributions $I_\theta(\mathbf{p})$
with weights
$n_\theta(\theta-\theta^{\parallel/\perp})=\exp(-\sin(\theta-\theta^{\parallel/\perp})^2/(2\sigma^2))$,
$\sigma^2=1-\cost$, and $\cost=0.9$.

Furthermore, to account for the rotation of the molecule about the polarisation axis for parallel
alignment we impose cylindrical symmetry of the photoelectron distribution about $Y$ by averaging
over $\phi$:
$\bar{I}_\phi(\mathbf{p})=(2\pi)^{-1}\int_{0}^{2\pi}\diff{\phi}\,R_\phi\left(I(\mathbf{p})\right)$
with the operator $R_\phi$ of rotation in the $X,Z$ plane. The final spectrum is obtained as
follows:
\begin{equation}
   \bar{I}^{\parallel/\perp}(p_X, p_Y) = \int\diff{p_z} \, \int\diff{\theta} \, n(\theta-\theta^{\parallel/\perp})
   \bar{I}_{\phi,\theta}(\mathbf{p})\,.
\end{equation}

To account for experimental-background in the simulations, a constant offset of $2\times10^{-8}$ was
added to the energy distributions, see \refsupp{fig:edf2}. For comparison, the spectra for perfectly
aligned configurations are reported in \refsupp{fig:edf3}. We point out that the background
correction shifts the numerically obtained cutoffs to lower energy, but does not affect the general
behaviour nor the difference of the cutoffs between the parallel and perpendicular configurations.

From the numerical simulations the crucial role of the usually neglected electron-electron
interaction for correctly describing the cutoff region in the parallel configuration became evident.
\subrefsupp{fig:edf4}{a} shows the decomposed contributions of the Kohn-Sham HOMO and HOMO-1 orbitals,
which highlight their distinct contributions to two distinct cutoffs, which are strongly separated
in intensity. In particular, the faint $10~U_p$ cutoff for the parallel case actually appears to be
uniquely determined by the HOMO-1, which does not have a node along the molecular axis, whereas
contributions from the HOMO were strongly suppressed by the presence of a node along the molecular
axis, \ie, parallel to the laser-polarisation axis. Second, the independent particle simulation
obtained by propagating the system with the Hartree, exchange, and correlation potentials frozen,
mimicking the widely used single-active electron model, presents a qualitatively different picture,
see \subrefsupp{fig:edf4}{b}. In particular the contribution of the HOMO-1 is highly overestimated and
for the parallel alignment the $10~U_p$ cutoff is restored, in clear contradiction with the
experiment. These results also confirm the importance of the coherent interaction between different
orbitals in strong-field ionisation~\cite{Akagi:Science325:1364}.

\section{Appendix~B: Semiclassical trajectory simulations}
\label{sec:semiclass_simulations}
Semiclassical trajectory simulations were carried out employing a simplified MO-ADK
(molecular-orbital Ammosov-Delone-Kra\v{\i}nov) approach, to create the initial wavepacket, in
conjunction with a classical continuum propagation in the combined laser-electric and Coulomb field
of the cation.

\subsection{Initial electron wavepacket}
In general, the phase-space distribution of the initial electron wavepacket was created in a similar
fashion as described before~\cite{Holmegaard:NatPhys6:428}. The ionisation probability in dependence
of the instantaneous electric field was obtained through the quasistatic ADK tunnelling theory
supplemented by an empirical extension to the barrier-suppression regime~\cite{Tong:JPB38:2593}.
Electric-field dependent ionisation potentials, $I_p(\vec{\epsilon})$, were computed through
second-order perturbation theory
\begin{equation}
   I_p(\vec{\epsilon}) = I_p^{(0)} - \Delta \vec{\mu} \!\cdot\! \vec{\epsilon} - \frac{1}{2}
   \vec{\epsilon}^\text{ T}\! \Delta \alpha \, \vec{\epsilon} \, ,
\end{equation}
assuming that ionisation occurs exclusively from the HOMO, as supported by the TDDFT results.
$I_p^{(0)}$ is the field-free ionisation potential, $\Delta\vec{\mu}$ and $\Delta\alpha$ are the
differences of dipole moment and polarizability tensor between cationic and neutral species,
respectively. Here, the measured field-free ionisation potential of
$I_p^{(0)}=11.19$~eV~\cite{Wang:JESRP47:167} was used in combination with calculated neutral and
cationic dipole moments and polarizabilities~\cite{Holmegaard:NatPhys6:428}. In \refsupp{fig:edf5}
the resulting time-of-birth distribution of a typical electron wavepacket is shown.

The classical tunnel exit was composed as
\mbox{$\vec{r}_0=-\vec{\epsilon}\,I_p/\epsilon^2$}~\cite{Bian:PRA84:043420}. The initial momentum
distribution in the plane transverse to the electric field vector at the instance of ionisation was
modelled according to the atomic ADK tunnelling theory with additional imprint of the initial
electronic state's nodal structure. The initial momentum component along the electric field vector
was obtained through nonadiabatic tunnelling theory~\cite{Li:PRA93:013402}. For the
parallel-alignment case the nodal line of the HOMOs along the molecular axis imprints onto the
momentum distribution at birth. Accordingly, the initial transverse momentum distribution for the
parallel alignment case was described as
\begin{equation}
   \omega_\parallel \!\left( p_{0,x}, p_{0,z} \right) \propto \left( p_{0,x}^2 + p_{0,z}^2 \right)
   \cdot
   \text{e}^{-\frac{\sqrt{2I_p}}{\epsilon}(p_{0,x}^2 + p_{0,z}^2)} \, .
\end{equation}
For the perpendicular alignment case the nodal plane orthogonal to the molecular axis imprints onto
the initial momentum distribution. Since it splits the HOMO's electron density unequally with a
ratio of 85:15~\cite{Holmegaard:NatPhys6:428}, its imprint was described as a nodal plane with a
damped peculiarity
\begin{equation}
	\beta = 1 - \frac{|\varrho_+ - \varrho_-|}{\varrho_+ + \varrho_-} \, ,
\end{equation}
with $\varrho_\pm$ representing the integral electron densities on the two sides of the nodal plane.
Hence, the distribution of initial transverse momenta for the perpendicular alignment case was set
up as
\begin{equation}
	\omega_\bot \!\left( p_{0,x}, p_{0,z} \right) \propto \left| p_{0,z} \right|^{2 \beta} \cdot
	\text{e}^{-\frac{\sqrt{2I_p}}{\epsilon}(p_{0,x}^2 + p_{0,z}^2)} \, .
\end{equation}
\refsupp{fig:edf6} illustrates exemplary transverse momentum distributions for both alignment cases.

\subsection{Classical propagation}
The electron wavepacket at birth into the continuum was sampled from the initial phase-space
distribution through rejection sampling, expanded as a coherent superposition of partial plane waves
and, subsequently, propagated in the combined electric laser and the cation's Coulomb field. The
singly charged cation was represented by a point charge $+e$.

The asymptotic electron phase after exposure to the combined electric field of laser and point
charge was obtained through~\cite{Shvetsov-Shilovski:PRA94:013415}
\begin{equation}
   \phi_\infty = -\vec{p}_0 \!\cdot\! \vec{r}_0 + I_p \!\cdot\! t_0 -\int_{t_0}^{\infty} \!\!
   \left( \frac{p^2}{2} - \frac{2}{r} \right) \text{d}t \, .
\end{equation}

At short distances between the electron and the point charge, $r<r_{2b}$, the laser-electric field
becomes negligible and the description of the electron motion reduces to a two-body problem. Here,
the threshold distance, $r_{2b}$, was chosen such that the corresponding Coulomb field is 1000 times
larger than the peak electric field of the laser. Accordingly, for the experimental peak electric
field of $\epsilon_0\approx0.048$~\au this threshold distance becomes
$r_{2b}=1/\sqrt{1000\,\epsilon_0}\approx0.14$~\au. The motion of the electron within the spherical
volume of radius $r_{2b}$ around the point charge could then be described conveniently as a Kepler
orbit. This approximation allows for direct computation of the electron's properties at its
symmetric exit point from the sphere: Its position and momentum vector at exit as well as its time
of flight between entry and exit of this sphere can be computed fully analytically. The phase
accumulated during its passage through the sphere is accessible through low-effort numerical
integration. In \refsupp{fig:edf7} a typical trajectory of an electron is shown as it performs a
swing-by around the point charge.

\bibliography{string,cmi}

\onecolumngrid%
\vspace*{0em}\vfill

\section*{Supplementary Figures}
\renewcommand{\figurename}{\textbf{Supplementary Figure}}
\renewcommand{\theHfigure}{SF.\thefigure}
\setcounter{figure}{0}
\begin{figure*}[b]
   \centering%
   \includegraphics[width=0.66\linewidth]{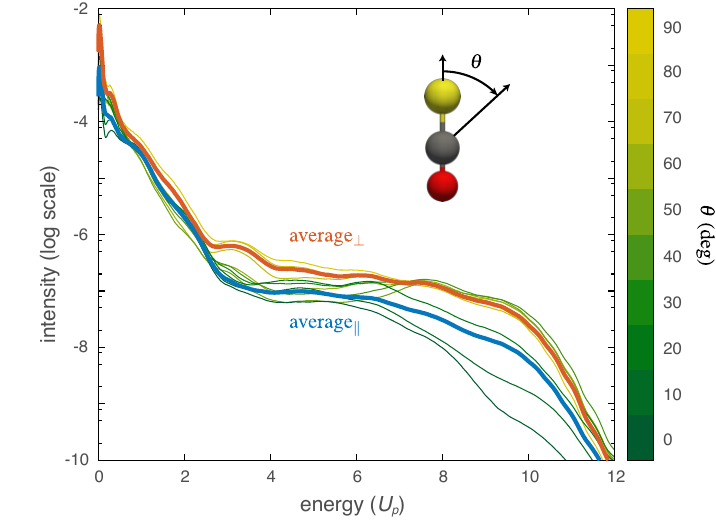}
   \caption{Simulated photoelectron spectra as a function of the angle $\theta$ between the laser
      polarisation and the molecular axis. The averaged spectra for a parallel and perpendicular
      alignment with a degree of alignment of $\cost=0.9$ are reported with thick lines.}
   \label{fig:edf1}
\end{figure*}

\begin{figure*}
   \centering%
   \includegraphics[width=0.75\linewidth]{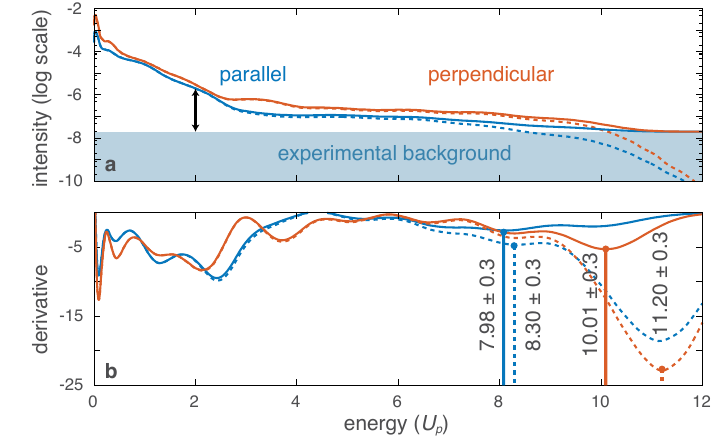}
   \caption{Effect of the experimental-background in the simulated photoelectron spectra.
      \textbf{a}: Simulated photoelectron distributions, for parallel and perpendicular alignment,
      averaged over $\theta$ according to the degree of alignment in the experiment
      $\bar{I}_{\parallel/\perp}(E)$ (dashed lines). A constant, $2\times10^{-8}$, is added to the
      simulated data to account for the experimental-background level (solid lines), \ie, to match
      the experimental contrast from $2$ to $12~U_p$. \textbf{b}: Derivatives of the photoelectron
      distributions. Regardless of the presence of the background the cutoffs for the parallel and
      perpendicular configurations present an observable difference which is in qualitative
      agreement with the results of Fig.2 of the main manuscript.}
   \label{fig:edf2}
\end{figure*}

\begin{figure*}
   \centering%
   \includegraphics[width=0.75\linewidth]{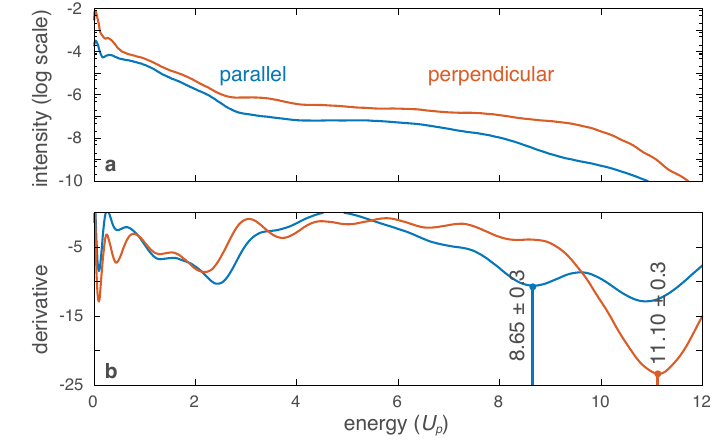}
   \caption{Same as \refsupp{fig:edf2} for perfect alignment and no experimental-background
      correction.}
   \label{fig:edf3}
\end{figure*}

\begin{figure*}
   \centering%
   \includegraphics[width=\linewidth]{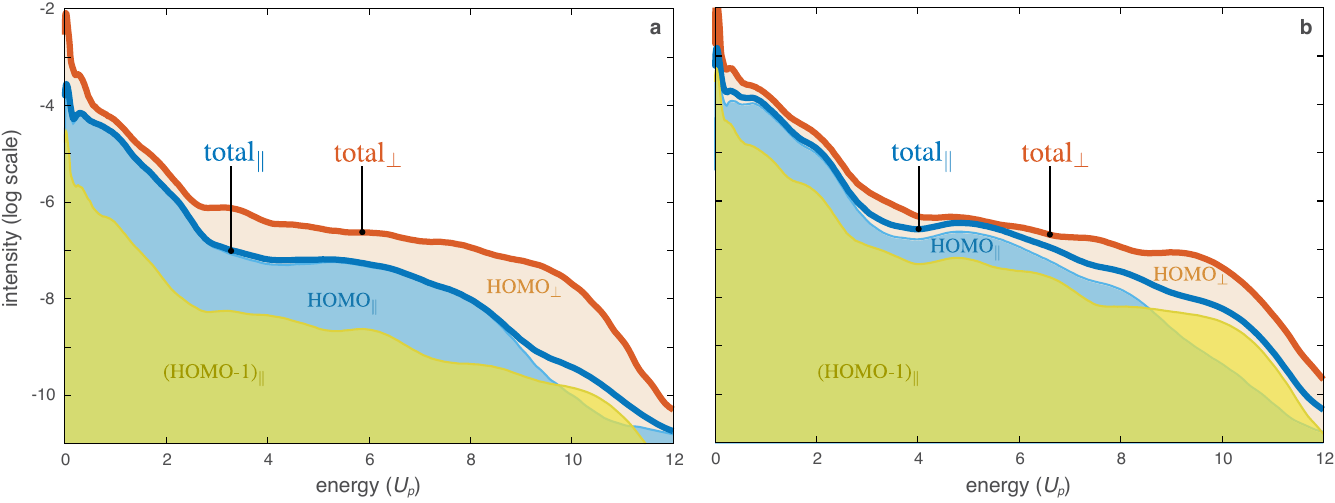}
   \caption{Role of electron-electron interaction and orbital contribution to the total
      photoelectron spectrum for perfect alignment. \textbf{a}: Spectra obtained with the
      fully-interacting-electrons TDDFT equations. \textbf{b}: Spectra for the
      non-interacting-electrons simulation obtained by freezing the Hartree exchange and correlation
      potentials to the neutral ground state potentials.}
   \label{fig:edf4}
\end{figure*}

\begin{figure*}
   \centering%
   \includegraphics[width=0.7\linewidth]{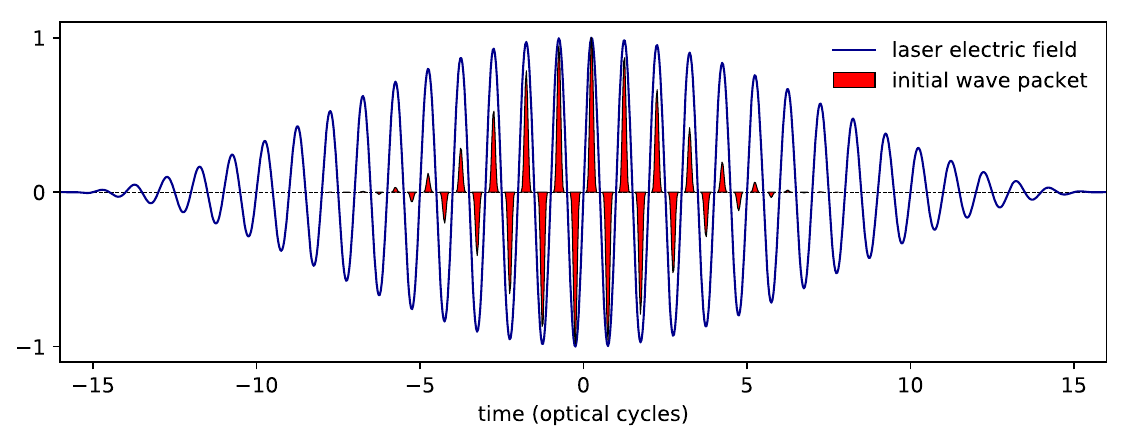}
   \caption{Time-of-birth distribution of an electron wavepacket for OCS at perpendicular alignment
      as obtained from the quasistatic ADK tunnelling theory. The distribution shown was normalised
      with respect to its maximum and multiplied with the sign of the instantaneous electric field.
      The electric field is displayed in units of $\epsilon_0$.}
   \label{fig:edf5}
\end{figure*}

\begin{figure*}
   \centering%
   \includegraphics[width=0.7\linewidth]{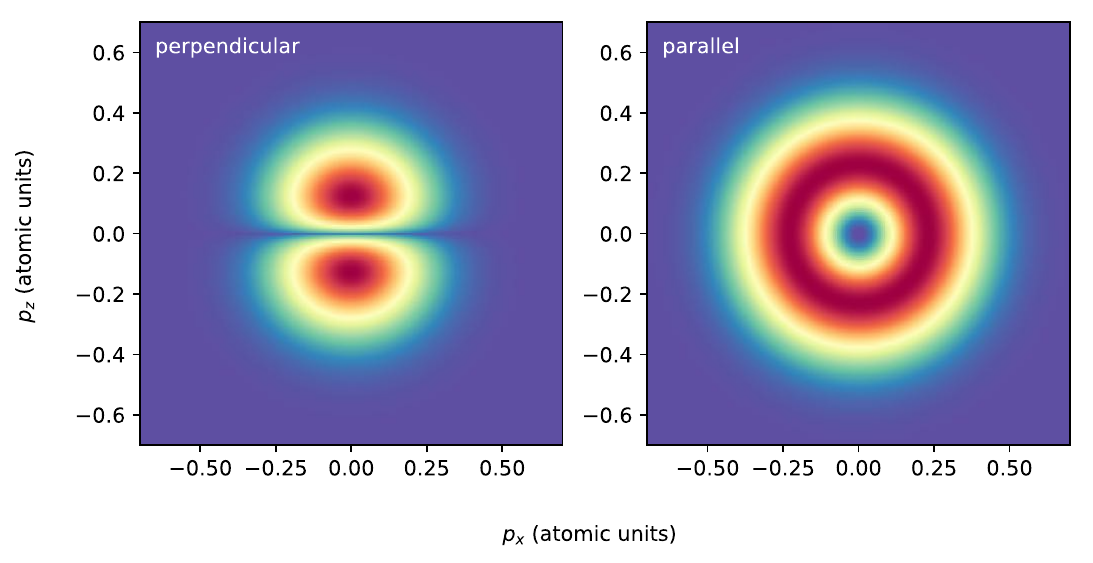}
   \caption{Transverse initial momentum distributions assuming the field-free ionisation potential
      and peak electric field.}
   \label{fig:edf6}
\end{figure*}

\begin{figure*}
   \centering%
   \includegraphics[width=0.5\linewidth]{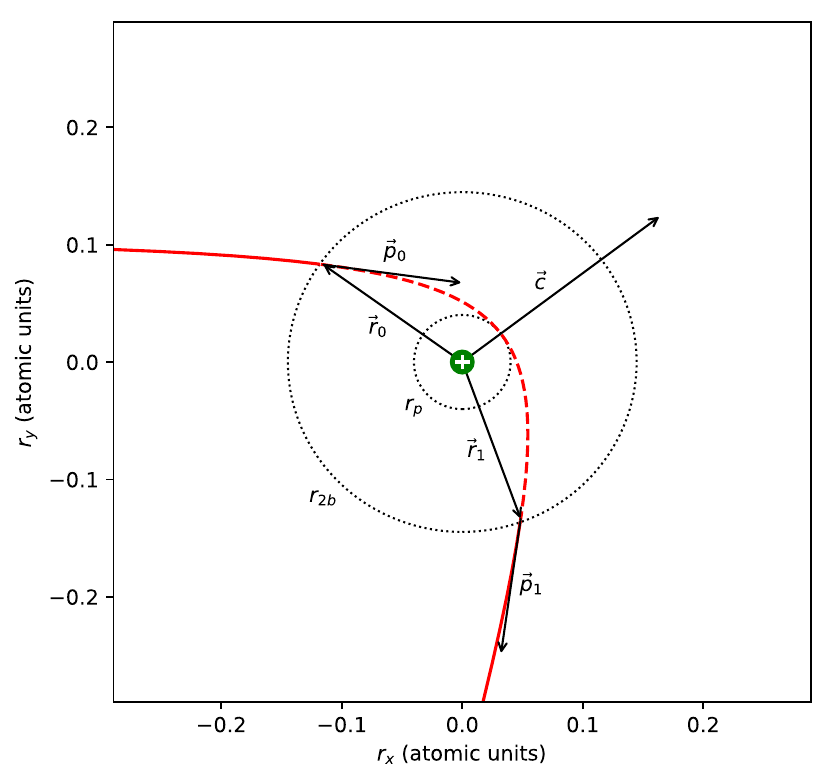}
   \caption{Example trajectory in the combined laser-electric and Coulomb field of a point charge
      $+e$ (green dot), numerically propagated for $r>r_{2b}$ (solid red line) and analytically
      approximated for $r\leq{}r_{2b}$ by means of a pure two-body interaction (dashed red line). By
      deduction of the orbital centre $\vec{c}$ and the distance at closest approach, the periapsis
      $r_p$, the symmetric exit position $\vec{r}_1$, the corresponding momentum vector $\vec{p}_1$,
      and the time spent within the sphere can be obtained analytically.}
   \label{fig:edf7}
\end{figure*}

\end{document}